\documentclass[reprint, amsmath,amssymb,aps]{revtex4-2}

\usepackage{graphicx,comment}
\usepackage{dcolumn}
\usepackage{bm,color,xcolor}

\DeclareMathOperator{\deform}{\gamma}

\begin{document}

\title{An intertwining between conformal dualities and ordinary dualities}
\author{Antonio Amariti}
  \email{antonio.amariti@mi.infn.it}
\affiliation{INFN, Sezione di Milano, Via Celoria 16, I-20133 Milano, Italy.}
\author{Simone Rota}
 \email{simone.rota@mi.infn.it}
\affiliation{Dipartimento di Fisica, Universit\`a degli Studi di Milano, Via Celoria 16, I-20133 Milano, Italy}

\date{\today}

\begin{abstract}
We discuss and reinterpret a 4d conformal triality recently discovered in the literature in terms of ordinary Seiberg duality.
We observe that a non-abelian global symmetry is explicitly realized by only two out of the three phase.
We corroborate the result by matching the superconformal index in terms of an expansion on the fugacities.
\end{abstract}

\maketitle

\section{Introduction}

The space spanned by the exactly marginal deformations, denoted as conformal manifold, is a powerful tool to explore the strongly coupled dynamics of a SCFT \cite{Leigh:1995ep,Kol:2002zt,Benvenuti:2005wi,Green:2010da,Kol:2010ub}.
Recently a classification program of 4d $\mathcal{N}=1$ gauge theories admitting a conformal manifold passing through weak coupling  has been 
developed in \cite{Razamat:2019vfd,Razamat:2020pra}.
This is a starting point for finding conformal dualities among, non-necessarily weakly coupled, SCFTs.

Many lagrangian descriptions of otherwise strongly coupled SCFT have been indeed obtained 
\cite{Razamat:2020pra,Razamat:2022gpm}.
Furthermore, new dualities between SCFTs admitting both a lagrangian description have been obtained
\cite{Zafrir:2019hps,Razamat:2020gcc,Zafrir:2020epd}. 
Such conformal dualities are interesting because they connect theories
that are not in general guaranteed to be connected by other IR dualities, e.g. Seiberg duality and its generalizations.

Furthermore one can go beyond and search for n-alities relating a higher number of models. For example in \cite{Razamat:2020pra} a conformal triality 
was claimed between an $SU(2)^7$ quiver gauge theory and other two models involving also  real gauge groups and tensorial matter.
In this triality the only symmetry preserved on the conformal manifold is $U(1)_R$. The central charges and the superconformal index have been shown to match.
It was also shown that the conformal manifolds have the same dimensions even if the mapping between the  exactly marginal couplings in each triality frame was not discussed.

Motivated by this last open problem here we prove that this conformal triality can obtained by  opportune chains of Seiberg \cite{Seiberg:1994pq} and Intriligator-Pouliot \cite{Intriligator:1995ne} dualities.
This result offers a natural mapping between the operators in the various phases.

 Furthermore we  turn off opportune superpotential couplings, increasing the amount of global symmetry. We show that this procedure does not spoil the triality even if only in some of the phases a non abelian global symmetry is explicitly realized by the 
action. We corroborate our interpretation by calculating the superconformal index and  studying its behavior between the phases with the enhanced global symmetry. 

We find some slices of the conformal manifold connecting pairs of weakly coupled cusps along which some discrete symmetries are preserved and we check this by charging the superconformal index with respect to these symmetries. These discrete symmetries become the Weyl subgroups of the non-abelian global symmetries in the loci where we have symmetry enhancement, for example at the weakly coupled cusps. We briefly comment on this behavior and make similar considerations about $\mathcal{N}=4$ SYM with Leigh-Strassler $\deform$-deformation \cite{Leigh:1995ep}.

\section{The conformal triality}

In this section we review the conformal triality found in \cite{Razamat:2020pra}. This triality  connects three frames with a
conformal manifold passing through weak coupling. By keeping the notation of \cite{Razamat:2020pra}
these three frames are  realized  by the following field theories 
\\
\\
{\bf Frame A}. An $USp(6)$ gauge theory with six totally antisymmetric tensors $\hat A_{i=2,\dots,7}$ and superpotential
\begin{eqnarray}
\label{WA}
W = \!\!\!\sum_{2 \leq i < j < k\leq 7}  \!\! \!\!  \text{Tr} \hat A_i \hat A_j \hat A_k+ \sum_{i=2}^{7} \text{Pf} \hat A_i
\end{eqnarray}
{\bf Frame B}.  An $\prod_{i=1}^7 SU(2)^7$ quiver gauge theories with all the possible chiral fields $X_{i,j}$ with $i\neq j$ connecting $SU(2)_i$ to $SU(2)_j$
and superpotential
\begin{equation}
W = \sum_{1\leq i < j < k \leq 7 } X_{ij} X_{jk} X_{k i}
\end{equation}
{\bf Frame C}.  An $SU(4)_1 \times SU(2)_3 \times SU(2)_{6}$ quiver with six antisymmetric tensors and three pairs of 
bifundamentals of $SU(4)_1 \times SU(2)_3$ and three antibifundamentals of $SU(4)_1 \times SU(2)_{6}$
and superpotential 
\begin{eqnarray}
\label{WC}
W\!&\!=\!&\!
A_6 (t_{16}^2\!+\!w_{16} v_{16}) 
\!+\! \!
A_5 (w_{16}^2 \!+\!t_{16} v_{16})
\!+\! \!
A_7 (v_{16}^2\!+\! t_{16} w_{16}\!)
\nonumber \\
&+&(A_2+A_3+A_4) ( w_{16} v_{16}+t_{16} v_{16}+ t_{16} w_{16})
\nonumber \\
\!&\!+\!&\!
A_3 (t_{31}^2\!+\!w_{31} v_{31})\!+\! \! A_4 (v_{31}^2\!+\! w_{31} t_{31})\!+\! \! A_2 (w_{31}^2\!+\! v_{31} t_{31}\!)
\nonumber \\
&+&
 (A_6+A_5+A_7)(w_{31} v_{31}+v_{31} t_{31}+ w_{31} t_{31})
\end{eqnarray}

Observe that for the ease of the reader above we have fixed the labels of the gauge groups compatibly 
with the ones obtained after the chains of dualities that we will study below.

The three models have the same central charges $a$ and $c$ and they only have 
an $U(1)_R$ global symmetry, that implies that there are no other 't Hooft anomalies to match. 
The conformal manifold has dimension dim$\mathcal{M}_c=21$
in all the three phases and the superconformal index has been shown to match at very large order in the fugacities.

Throughout this paper the color indices of all the fields are omitted and the contractions between them are understood.

\section{Conformal triality from Seiberg duality}
\label{confromseiberg}
In this section we show that the three models introduced above can be mapped through chains of Seiberg and Intriligator-Pouliot dualities. 
This idea is similar to the one used to show the enhancement of $A_7$ to $E_7$ for $USp(2)$ SQCD with eight fundamentals \cite{Dimofte:2012pd,Razamat:2017hda}.
Indeed one can consider either a gauge group as an $SU(2)$ or as $USp(2)$, and dualize such node with a different interpretation for the flavor symmetry.
Indeed if the gauge group is considered as $SU(2)$ one needs to distinguish fundamental and antifundamental representations, and it reflects in the interpretation of the mesonic and baryonic deformations. On the other hand if a gauge group is considered as $USp(2)$ only mesons are available. 
Such a different interpretation becomes relevant after the duality, because if the number of $USp(2)$ fundamentals is higher than eight then the rank of the gauge group is $SU(n>2)$ or $USp(2n>2)$. Furthermore the superpotential deformations have a different role in the dual phase because, as anticipated, they can involve either mesons or baryons.
Another consequence of our derivations is that the matching of the integrals representing the superconformal index and the operator mapping on the conformal manifold follows automatically from the duality.

\subsection{From  Frame B to  Frame A}
\label{pippo}

In this case we start by by treating the $SU(2)_1$ gauge node as $USp(2)$.
The node has $2N_f=12$ fundamental and it is dual to an $USp(6)$ gauge group. The mesons associated to this gauge group are of two types. There are 
mesons  in the form $M_{ij} \equiv X_{i1} X_{j1}$, in the bifundamental of $SU(2)_i \times SU(2)_j$
with $i \neq j$ and mesons $S_i = X_{i1}^2$, that correspond to singlets of $SU(2)_i$.
The final superpotential is
\begin{eqnarray}
W &=& \sum_{2\leq i<j \leq 7} M_{ij} X_{ij} + \sum_{i \leq j} M_{ij} q_{1i}q_{1j} 
\nonumber \\
&+&
 \sum_{i=2}^{7} S_i q_{1i} q_{1i} + \sum_{2 \leq i < j < k\leq 7} X_{ij}X_{jk}X_{ki}
\end{eqnarray}
where $q_{1i}$ are the dual bifundamentals charged under $USp(6)_1 \times SU(2)_i$.
By integrating out the massive fields one is left with 
\begin{eqnarray}
W = \sum_{i=2}^{7} S_i q_{1i} q_{1i} + \sum_{2 \leq i < j < k\leq 7} (q_{1i}q_{1j}q_{1k})^2
\end{eqnarray}
Next we can dualize each $SU(2)_{i=2,\dots,7}$ by treating them as $USp(2)$ nodes, each one with six fundamentals.
Each $USp(2)$ factor gives rise to 15 singlets, corresponding to six antisymmetric $A_{2,\dots,7}$ tensors of $USp(6)$.
The superpotential is
\begin{eqnarray}
W = \sum_{i=2}^{7} S_i A_{i}^{(0)} +\!\!\! \sum_{2 \leq i < j < k\leq 7}  \!\! \!\! \text{Tr} A_i A_j A_k + \sum_{i=2}^{7} \text{Pf} A_i
\end{eqnarray}
The F-terms of $S_i$ set $A_i^{(0)}$ to zero, and we are left with six totally antisymmetric tensors $\hat A_i$ (i.e. of  dimension 14) interacting through the superpotential (\ref{WA}).

\subsection{From  Frame B to  Frame C}
\label{pluto}
\begin{figure}
\begin{center}
\includegraphics[width=6cm]{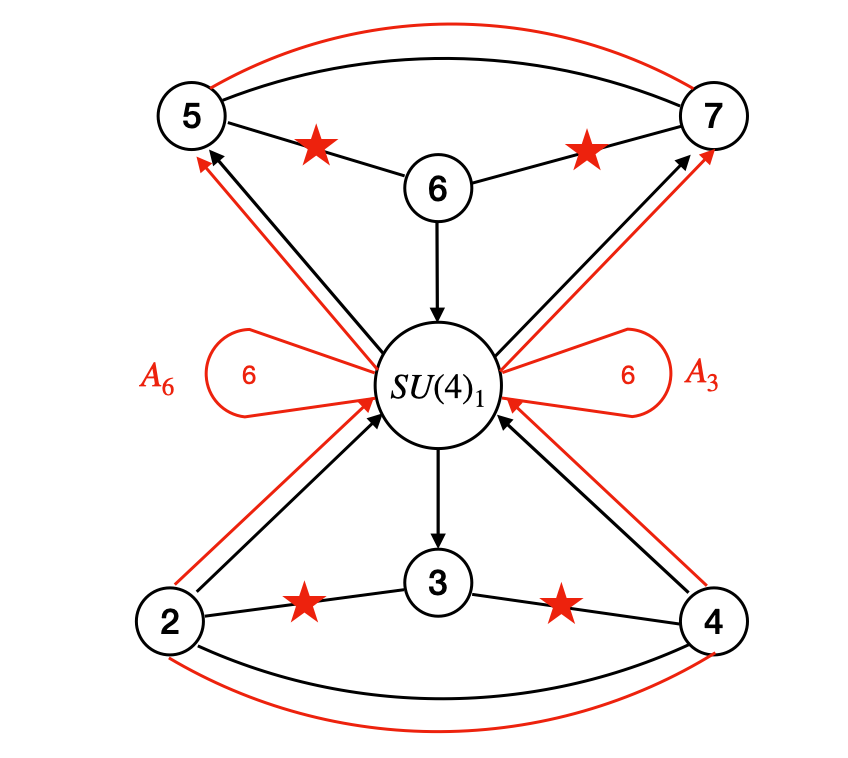}
\caption{Quiver obtained after Seiberg duality on $SU(2)_1$ and Intriligator Pouliot on $SU(2)_3$ and $SU(2)_6$. The bifundamentals between the nodes 2,4 and 5,7 are massive and are integrated out.}
\label{Quiveb33}
\end{center}
\end{figure}

\begin{figure}
\begin{center}
\includegraphics[width=8cm]{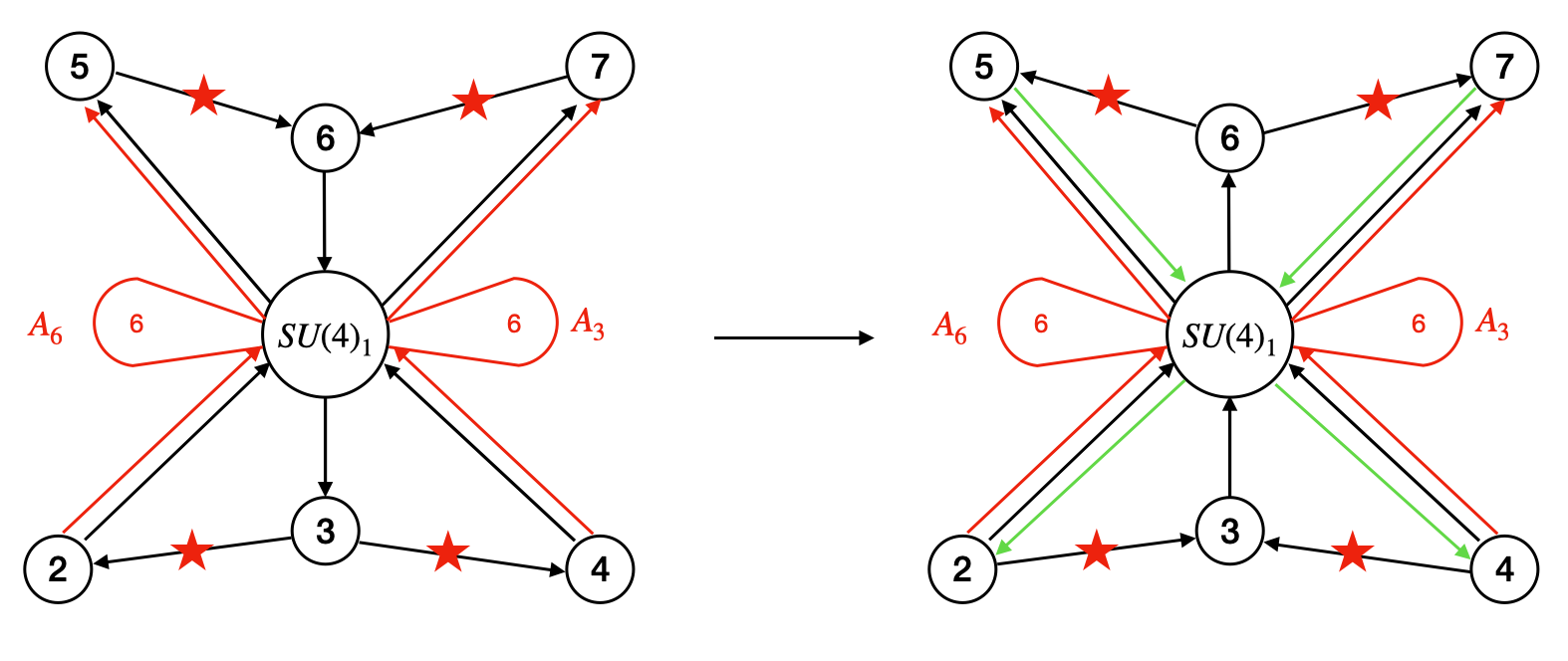}
\caption{Quiver obtained from the one of Figure \ref{Quiveb33} by Seiberg duality on $SU(2)_3$ and $SU(2)_6$. The orientation of the arrows connecting $SU(2)_5$ with $SU(2)_{4,6}$ and  $SU(2)_3$ with $SU(2)_{2,4}$ is necessary in order to construct the mesons of the dual phase.}
\label{Quiveb44}
\end{center}
\end{figure}

In the second case we consider $SU(2)_1$ as unitary, i.e. we apply the ordinary Seiberg duality.
This implies that we have to make a choice on the $SU(2)$ representations, dividing them into fundamentals and antifundamentals. The dual gauge group is then $SU(4)$ and we must pay attention to the structure of the dual superpotentials, because of the presence of interactions involving the baryonic deformations that have to be threaten with
some care.
We proceed by splitting the index $I=1,\dots,7$ as $I=\{1,i=2,3,4,\alpha=5,6,7\}$ and the fields are then
$X_{1i}$, $X_{\alpha1}$, $X_{ij} = X_{ji}$ and  $X_{\alpha \beta} = X_{\beta \alpha}$.
The superpotential is 
\begin{eqnarray}
W &=& X_{23} X_{34} X_{42}+X_{56}X_{67}X_{75}+\sum_{2\leq i<j\leq 4} \sum_{\alpha=5}^7 X_{ij} X_{j\alpha}X_{\alpha j}
\nonumber \\
&+&
 \sum_{5\leq \alpha<\beta\leq 7} \sum_{i=2}^4 X_{i\alpha} X_{\alpha \beta }X_{\beta i}
+
\sum_{i=2}^4 \sum_{\alpha=5}^7  X_{1i} X_{i\alpha}X_{\alpha 1}
\nonumber \\
&+&
\sum_{2\leq i<j\leq 4} X_{1 i} X_{1 j} X_{i j}+\sum_{5\leq \alpha < \beta \leq 7}X_{\alpha 1} X_{\beta 1} X_{\alpha \beta
}
\end{eqnarray}
The first four terms are spectator of the duality, while the others  involve the mesonic deformations (first term) and the baryonic ones (last two terms).
The dual superpotential, after integrating out the massive fields, becomes
\begin{eqnarray}
W &=& 
X_{23} X_{34} X_{42}+\sum_{i=2}^4 \sum_{5\leq \alpha < \beta \leq 7} q_{i1}^2 q_{1\alpha} q_{1 \beta}  X_{\alpha \beta}
\nonumber \\
&+&
(\prod_{i=2}^{4} q_{i1}) (\sum_{i\neq j \neq k} q_{i1} X_{jk})
 \\
&+&
X_{56}X_{67}X_{75}+\sum_{\alpha=5}^7 \sum_{2\leq i <j\leq 4} q_{1\alpha}^2 q_{i1}q_{j1}X_{ij} +
\nonumber \\
&+&
(\prod_{\alpha=5}^{7} q_{1 \alpha}) (\sum_{\alpha\neq \beta \neq \gamma} q_{1\alpha} X_{\beta \gamma})\nonumber 
\end{eqnarray}
We then proceed by applying a Intriligator Pouliot duality on $SU(2)_3=USp(2)_3$ and $SU(2)_6=USp(2)_6$.
The quiver becomes the one in the Figure \ref{Quiveb33}
where we highlighted in red the mesons of the dualities.
The dual superpotential, after integrating out the massive fields, becomes

\begin{eqnarray}
W
&=&
p_{32}p_{34} ({\color{red}A_3} q_{21} q_{41}\!+\! {\color{red}A_6} q_{21} q_{41}\!+\!q_{21} q_{41} q_{15}^2\!+\!q_{17}^2 q_{21} q_{41})
\nonumber \\
&+&
p_{56} p_{76}  ({\color{red}A_3} q_{15} q_{17}\!+\! {\color{red}A_6} q_{15} q_{17}\!+\!q_{15} q_{17} q_{21}^2\!+\!q_{15} q_{17} q_{41}^2)
\nonumber \\
&+&
{\color{red}s_2} p_{32}^2+{\color{red}s_4} p_{34}^2+{\color{red}s_5} p_{56}^2+{\color{red}s_7} p_{76}^2 + {\color{red}A_3} p_{13}^2+{\color{red}A_6} p_{61}^2
\nonumber \\
&+&
{\color{red}N_{21}} p_{13} p_{32} + {\color{red}N_{41}} p_{13}p_{34}
+
{\color{red}N_{15}} p_{56} p_{61}+ {\color{red}N_{17}} p_{76} p_{61}  
\nonumber \\
&+&
{\color{red}A_6} \left({\color{red}N_{21}} q_{21}+{\color{red}N_{41}} q_{41}\right)
+
{\color{red}A_3} \left({\color{red}N_{15}} q_{15}+{\color{red}N_{17}} q_{17}\right)
\nonumber \\
&+&
{\color{red}N_{15}} \left(q_{15} q_{17}^2+q_{15} q_{21}^2+q_{15} q_{41}^2\right)
\nonumber \\
&+&
{\color{red}N_{41}} \left(q_{41} q_{15}^2+q_{17}^2 q_{41}+q_{21}^2 q_{41}\right)
\nonumber \\
&+&
{\color{red}N_{17}} \left(q_{17} q_{15}^2+q_{17} q_{21}^2+q_{17} q_{41}^2\right)
\nonumber \\
&+&
{\color{red}N_{21}} \left(q_{21} q_{15}^2+q_{21} q_{41}^2+q_{17}^2 q_{21}\right)
\end{eqnarray}
The next step consists of acting with two Seiberg dualities on the nodes $SU(2)_3$ and $SU(2)_6$. It requires a choice on the fundamentals and anti-fundamentals (see Figure \ref{Quiveb44}). After the duality we have represented the dual quiver by highlighting the mesons of this step in green.
 The dual superpotential, after  integrating out the massive fields,  becomes
\begin{eqnarray}
\label{Walmost}
W
&=&
{\color{red} s_2} t_{23}^2+{\color{red} s_4} t_{43}^2+
{\color{red} s_5}
 t_{65}^2+{\color{red} s_7}  t_{67}^2+{\color{red} A_3} t_{31}^2\!+\!{\color{red} A_6} t_{16}^2
\nonumber \\
&+&
t_{23} t_{43} \left({\color{red} A_3}  q_{21} q_{41}+{\color{red} A_6}  q_{21} q_{41}\!+\!q_{21} q_{41} q_{15}^2\!+\!q_{17}^2 q_{21} q_{41}\right)
\nonumber \\
&+&
t_{65} t_{67}  \left({\color{red} A_3}  q_{15} q_{17}+{\color{red} A_6} q_{15} q_{17}\!+\!q_{15} q_{17} q_{21}^2\!+\!q_{15} q_{1,7} q_{4,1}^2\right)
\nonumber \\
&+&
t_{16} t_{65}  \left(q_{15} q_{17}^2+q_{15} q_{21}^2+q_{15} q_{41}^2+{\color{red} A_3} q_{15} \right)
\nonumber \\
&+&
 t_{43} t_{31}  \left(q_{41} q_{15}^2+q_{17}^2 q_{41}+q_{21}^2 q_{41}+{\color{red} A_6} q_{41}\right)
\nonumber \\
&+&
 t_{16}t_{67}  \left(q_{17} q_{15}^2+q_{17} q_{21}^2+q_{17} q_{41}^2+{\color{red} A_3}  q_{17} \right)
\nonumber \\
&+&
 t_{23} t_{31} \left(q_{21} q_{15}^2+q_{21} q_{41}^2+q_{17}^2 q_{21} +{\color{red} A_6} q_{21} \right)
\nonumber 
\end{eqnarray}

The final dualities  that we need to perform are Intriligator-Pouliot dualities on nodes $2,4,5,7$.
These four gauge  nodes are s-confining and we are left with the confined degrees of freedom consisting of the following 
fields 
\begin{eqnarray}
\label{dm1}
\mathcal{M}_{2} 
&=&
\left( \begin{array}{cc}
	q_{21}^2  & t_{23} q_{21} \\
	(t_{23} q_{21})^T & t_{23}^2 
	\end{array} \right)
	=
	\left( \begin{array}{cc}
	A_2 & w_{31} \\
	w_{31}^T & r_2
	\end{array} \right)
\end{eqnarray}
\begin{eqnarray}
\label{dm2}
\mathcal{M}_{4} 
&=&
\left( \begin{array}{cc}
	q_{41}^2  & t_{43} q_{41} \\
	(t_{43} q_{41})^T & t_{43}^2 
	\end{array} \right)
	=
	\left( \begin{array}{cc}
	A_4 & v_{31} \\
	v_{31}^T & r_4
	\end{array} \right)
\end{eqnarray}
\begin{eqnarray}
\label{dm3}
\mathcal{M}_{5} 
&=&
\left( \begin{array}{cc}
	q_{15}^2  & t_{65} q_{15}  \\
	( t_{65} q_{15})^T & t_{65}^2 
	\end{array} \right)
	=
	\left( \begin{array}{cc}
	A_5 & w_{16} \\
	w_{16}^T & r_5
	\end{array} \right)
\end{eqnarray}
\begin{eqnarray}
\label{dm4}
\mathcal{M}_{7} 
&=&
\left( \begin{array}{cc}
	q_{17}^2  & t_{67} q_{17}  \\
	( t_{67} q_{17})^T & t_{67}^2 
	\end{array} \right)
	=
	\left( \begin{array}{cc}
	A_7 & v_{16} \\
	v_{16}^T & r_7
	\end{array} \right)
\end{eqnarray}
In this case we are left with an $SU(4)_1 \times SU(2)_3 \times SU(2)_6$ gauge theory.
The fields $A_{2,\dots,7}$ are in the antisymmetric of $USp(4)$,  the fields $t_{16}$
$v_{16}$ and $w_{16}$ are in the bifundamental of $SU(4)_1 \times SU(2)_6$ and the fields
 $t_{31}$
$v_{31}$ and $w_{21}$ are in the anti-bifundamental of $SU(2)_3 \times SU(4)_1$.
The fields $r_{3,\dots,6}$ and $s_{3,\dots,6}$ are gauge singlets. 
There are two types of  superpotential terms after the confinement.
One type of term corresponds to the contribution of the Pfaffians Pf$\mathcal{M}_{2,4,5,7}$, 
while the other terms are deformations
obtained by applying the duality map (\ref{dm1})--(\ref{dm4}) to (\ref{Walmost}).
After integrating out the massive fields we are left with (\ref{WC}).

\section{One step further: turning off $W$}

In this section we go beyond the results obtained above and we turn off 
 some of the superpotential couplings in the $SU(2)^7$ model.
We then follow the chain of dualities studied above.
We distinguish two cases, in the first case we keep only the  superpotential 
term
\begin{equation}	\label{eq:WI}
W_{B} = \sum_{2 \leq < i < j\leq 7} X_{1j}X_{1j}X_{ij}
\end{equation}
as non vanishing, while we turn off the terms $X_{ij}X_{jk}X_{ki}$ with 
$2 \leq i < j < k\leq 7 $.
In the second case keep only 
\begin{eqnarray}
W_{B} = X_{23} X_{34} X_{42}+X_{56}X_{67}X_{75}+\sum_{i=2}^4 \sum_{\alpha=5}^7  X_{1i} X_{i\alpha}X_{\alpha 1}\nonumber 
\\ \label{eq:WII}
\end{eqnarray}
Repeating the derivation above we start dualizing in the first case  as done in Subsection  \ref{pippo} and in the second case 
as done in Subsection \ref{pluto}.
In the first case we obtain the $USp(6)$ gauge theory with six totally antisymmetric tensors and 
\begin{eqnarray}
\label{nptW}
W_{A}=\sum_{i=2}^{7} \text{Pf} \hat A_i
\end{eqnarray}
while in the second case we obtain the $SU(2)^2 \times SU(4)$ quiver, but this time with superpotential 
\begin{eqnarray}
W_{C}\!=\!
A_6 t_{16}^2 \!+\! A_5 w_{16}^2\!+\!A_4 v_{16}^2
\!+\!
A_3  t_{31}^2\!+\! A_2 w_{31}^2\!+\!A_4 v_{31}^2
\nonumber \\
\end{eqnarray}
Some comments are in order.
First, turning off the superpotential deformations in the $SU(2)^7$ model does not rescue any continuous  symmetry.
This can be observed also also by acting with Intriligator-Pouliot duality on $SU(2)_1$: the $SU(2)^6 \times USp(6)$ dual
quiver has an anomalous axial symmetry associated to each bifundamental.  
In the $USp(6)$ dual theory this symmetry breaking pattern is more subtle, because it is broken by the non-perturbative contribution to the superpotential (\ref{nptW}).

On the other hand a global symmetry enhancement is observed in the dual phase. Indeed by keeping only the superpotential \eqref{eq:WII} the final model has an emergent non abelian $SU(3)^2$ global symmetry.

This symmetry enhancement can be understood from the duality structure that we have performed. Indeed in the second step we have make the choice of dualizing the nodes $SU(2)_3$ and  $SU(2)_6$.
There are other symmetric choices, we could have either dualized one group between  $SU(2)_{2}$ and $SU(2)_{4}$ or
one group between $SU(2)_{5}$ and $SU(2)_{7}$. For each choice the final $SU(4)^2 \times SU(2)$ model would have been
completely analogous, reflecting a self duality. This self duality is actually the Weyl group of the $SU(3)^2$ global symmetry that emerges in the model.

We have then computed the superconformal index of the  model in {\bf Frame C}, by charging it with respect to the $SU(6)\times SU(3)^2$ non abelian global symmetry at the free point \cite{Rastelli:2016tbz,Spiridonov:2011hf}. By expanding in the fugacities we obtaine
\begin{equation}
\begin{split}
\mathcal{I}&_{\mathbf{C}}(pq,\mathbf{a},\mathbf{b},\mathbf{c})=
1 +
(pq)^{2/3} \mathbf{21}_a 
 \\ 
&+ (pq) \left(	-\mathbf{35}_a -\mathbf{8}_b - \mathbf{8}_c + \mathbf{6}_a (\mathbf{6}_b+\mathbf{6}_c)	\right) + \dots
\end{split}
\label{eq:IndexC}
\end{equation}
where $\mathbf{a}$ are the fugacities for the $SU(6)$ global symmetry
 and $\mathbf{b},\mathbf{c}$ are the fugacities for the $SU(3)^2$ global symmetries.
 We use conventions where the character of the fundamental representation of $SU(N)$ is $\mathbf{N}_d=d_1 + \dots + d_N$ and the fugacities $\mathbf{d}$ satisfy $\prod_{i=1}^N d_i = 1$.
 In the slice of the conformal manifold where only a diagonal $SU(3)^3$ subgroup of the global symmetry is preserved the superconformal index is obtained form the one above by identifying $a_i = b_i$ and $a_{i+3}=c_i$ for $i=1,2,3$.

A further step consists in looking for the presence of such a symmetry in {\bf Frame A}.
At the free point of this frame the global symmetry is $SU(6)$. The index charged under this symmetry is
\begin{eqnarray}
\mathcal{I}_{\mathbf{A}}(pq,\mathbf{t}) = 1 +
(pq)^{2/3} \mathbf{21}_t 
+
(pq) \left( - \mathbf{35}_t + \mathbf{56}_t
\right) + \dots
\nonumber\\
\label{eq:indexBA}
\end{eqnarray}
where $t_{1,\dots,6}$ are the fugacities associated with $SU(6)$.
In the slice of the conformal manifold where an $SU(3)^2$ global symmetry is preserved the superconformal
 index is obtained by imposing $t_3 = (t_1 t_2)^{-1}$ and $t_6=(t_4 t_5)^{-1}$. 
We observed that the two expression \eqref{eq:indexBA} and \eqref{eq:IndexC} match if we identify $t_i=a_i$ for $i=1,\dots,6$.

We can interpret this result as an identification of a sub-manifold of the conformal manifold preserving 
the $SU(3)^2$ symmetry. This corroborates the idea that also the theory in  {\bf Frame B} has such an enhancement when the superpotential \eqref{eq:WII} is turned on, even if such a symmetry is not emergent at lagrangian level. We regard this as an example of an accidental non-abelian symmetry. As we argued above the Weyl group of the symmetry is visible as permutations of the gauge nodes $2,3,4$ and $5,6,7$, while the Cartan subalgebra emerges in the IR.

Finally we consider the superconformal index for \textbf{Frame A} with superpotential \eqref{nptW}. The Weyl subgroup of the $SU(6)$ global symmetry is preserved while the Cartan is broken to $U(1)^5 \to (\mathbb{Z}_3)^6/\mathbb{Z}_3$. We can implement this in the superconformal index \eqref{eq:indexB} by imposing $t_i^3=1$. By looking for example at order $pq$ we find:
\begin{eqnarray}
&-& \mathbf{35}_t + \mathbf{56}_t =
-5 - \sum_{j>i}^{6} \frac{t_i}{t_j} + \sum_{i=1}^{6} t_i^6 + \sum_{i>j}^{6} t_i^2 t_j  
\nonumber
\\&+&\sum_{i>j>k}^{6} t_i t_j t_k =
1+\sum_{i>j>k}^{6} t_i t_j t_k = 1 + \mathbf{20}_t
\end{eqnarray}
where with an abuse of notation we organize the contributions to the superconformal index in characters of $SU(6)$ in order to have a more compact notation. We stress that the fugacities are associated to the discrete symmetry of the model and the index is meaningful in every point of the conformal manifold where this symmetry is preserved, not only at the free point where we actually have an $SU(6)$ global symmetry. The full index reads:
\begin{eqnarray}
\mathcal{I}_\mathbf{A} (pq,\mathbf{t}) = 1 + (pq)^{2/3} \mathbf{21}_t + (pq) (1 + \mathbf{20}_t) + \dots
\nonumber\\
\end{eqnarray}

This theory is dual to \textbf{Frame B} with superpotential \eqref{eq:WI}, which is consistent with a discrete symmetry $S_6$ exchanging the nodes $2,\dots,7$ and an $(\mathbb{Z}_3)^6/\mathbb{Z}_3$ acting as:
\begin{eqnarray}
	(\mathbb{Z}_3)_i&:& X_{i1} \to e^{\frac{\pi i}{3}} X_{i1} 
	\\
	&&X_{ij} \to e^{-\frac{\pi i}{3}} X_{ij}, \qquad j \neq 1 
\end{eqnarray}
We can charge the superconformal index with respect to the discrete global symmetry along the lines of   \cite{Spiridonov:2011hf} where the discrete $\mathbb{Z}_{2N_f}$ symmetry of $SO(N)$ SQCD was studied. We charge the index 
under the $ U(1)_i$ symmetries under which $X_{1i}$ has charge $-\frac{1}{2}$ and $X_{ij}$ with $j\neq1$ has charge $\frac{1}{2}$. These symmetries are anomalous and are broken to the $(\mathbb{Z}_3)^6/\mathbb{Z}_3$ discrete symmetry by instanton effects \cite{tHooft:1976rip} (see also \cite{Csaki:1997aw}), therefore we impose $w_i^3=1$ on the corresponding fugacities. We obtain:
\begin{eqnarray}
\mathcal{I}_\mathbf{B} (pq, \mathbf{w}) = 1 + (pq)^{2/3} \mathbf{21}_w  + (pq) (1+\mathbf{20}_w)+\dots
\nonumber\\
\label{eq:indexB}
\end{eqnarray}
The two indices match under the identification $w_i = t_i$. We checked that this is true up to order $(pq)^{3/2}$. We interpret this as a matching between the discrete global symmetries. As we already pointed out, in \textbf{Frame A} the $S_6$ discrete symmetry is the Weyl of the $SU(6)$ global symmetry at the free point, broken by the superpotential \eqref{nptW}. The full $SU(6)$ symmetry can be recovered by turning off the superpotential \eqref{nptW}, which corresponds to a 
marginal deformation parametrized by an operator uncharged under the $S_6$ discrete symmetry. From the superconformal index \eqref{eq:indexB} we see that there is only one such operator, therefore we claim that \textbf{Frame B} with the superpotential \eqref{eq:WI} and the (unique) additional exactly marginal deformation uncharged under $S_6$ is dual to \textbf{Frame A} at the free point, for a suitable value of the additional deformation. In this point the global symmetry enhances to $SU(6)$. The Weyl of this symmetry is visible in the UV as permutation of the nodes $2,\dots,7$ while the Cartan is accidental.

Similarly one can analize the behavior of the discrete symmetries between \textbf{Frame B} with superpotential \eqref{eq:WII} and \textbf{Frame C}, the procedure is analogous to the one carried out in this section and we will not describe it in this paper.

\section{Discussion}

In this note we have interpreted the conformal triality found in \cite{Razamat:2020pra} in terms of ordinary Seiberg and Intriligator-Pouliot duality.
We have first considered the $SU(2)^7$ quiver gauge theory with the whole set of superpotential marginal deformation turned on.
This corresponds to {\bf Frame B}.
This deformation explicitly breaks the global symmetry to $U(1)_R$. We have then dualized one of the $SU(2)$ gauge node by treating it as  $USp(2)$. The resulting quiver, corresponding to $USp(6) \times USp(2)^6$ with bifundamentals connecting $USp(6)$ with each $SU(2)$ factor, corresponds to an intermediate phase, where the various $USp(2)$ gauge nodes are actually confining gauge theories. In the IR such gauge groups can be traded with sets of antisymmetric mesons. Such antisymmetric mesons correspond to the antisymmetric tensors of the $USp(6)$ gauge group, that is treated as a spectator in this case.
By carefully tracing the superpotential deformations only the totally antisymmetric part of these fields survives in the IR.
This final model corresponds to  {\bf Frame A}. Analogously, starting from the model denoted as {\bf Frame B}  we have constructed the one denoted as  {\bf Frame C}, by iterating a series of Seiberg and Intriligator-Pouliot dualities.

As a further step we have shown that such dualities between the models persist if we turn off opportune superpotential deformations
in the $SU(2)^7$ quiver.
Actually in this case we have found an interesting picture when computing the superconformal index.
Indeed the visible global symmetries in this case is still $U(1)_R$ in  {\bf Frame B} and 
{\bf Frame A}, while it enhances to  $SU(3)^2 \times U(1)_R$ in {\bf Frame C} .

This symmetry breaking structure in {\bf Frame A} resembles the discussion of the mismatch of global symmetries in the Berkooz deconfiment of gauge charged two-index tensors \cite{Berkooz:1995km,Sudano:2011aa,Craig:2011tx}. Actually in the case of Berkooz deconfinement the 
situation is different, indeed in that case the claim is that the presence of non perturbative effects breaks the extra global symmetry in the deconfined phase. Here instead the non-perturbative effect break the non abelian global symmetry in the confined phase idenfied by {\bf Frame A}.
On the other hand we observed a symmetry enhancement in  {\bf Frame C}, because of a self duality emergent from a permutational $S_3^2$ symmetry.

This is an important issue that needs a clarification. On one hand the integral identities used to fully match the superconformal indices  between {\bf Frame B} and {\bf Frame C} holds  only when  $U(1)_R$ is the only global symmetry. This is because the constraints (usually referred as balancing conditions) necessary to match the superconformal index requires that any other global symmetry is absent.
Anyway we are still free to charge the superconformal index with respect to  $SU(3)^2$ in {\bf Frame C}, because this global symmetry is realized by the action.

By a direct computation we ave further shown the model in {\bf Frame A} can be deformed to reach a point in the conformal manifold 
where an $SU(6)$ global symmetry is realized. We have shown that this index can be matched with the one obtained from 
{\bf Frame C} once it is specialized on an $SU(3)^2$ fugacity.
This is compatible with the claim that the theories are conformally dual and signals the fact that along these symmetries one can move along weakly coupled directions in the conformal manifold preserving indeed the global symmetry structure.
One can also use the operator map inherited from the underlining Seiberg and Intriligator-Pouliot dualities in order to have an explicit parametrization of such sub-manifold.
On the other hand, for more general enhancements and for a matching with the model in {\bf Frame B} strongly coupled 
directions  have to be taken into account and the lagrangian description is not sufficient to have the whole picture.

It should be interesting to further investigate on such phenomenon, looking for other conformally dual models, of the type studied in \cite{Razamat:2019vfd,Razamat:2020pra,Razamat:2022gpm}. Finding examples of such dualities connected through ordinary Seiberg-like  dualities with possibly some gauge factor that locally confines. We have not found so far examples of this type in the literature. Nevertheless an useful starting point 
consists in finding models that are conformally dual to the ones discussed in \cite{Razamat:2020pra}, because in such cases the conformal manifold passes through weak coupling.  

Another interesting question that arises in this discussion is the possibility of observing the pattern of global symmetry enhancement
from the superconformal index along the lines of \cite{Spiridonov:2008zr,Dimofte:2012pd,Razamat:2017hda,Razamat:2017wsk,Razamat:2018gbu}. 
We have shown that, through a sequence of controlled motions on the conformal manifold and Seiberg-like dualities, some discrete subgroup of the global symmetry can be mapped between two weakly coupled frames.  For example we find a one-dimensional slice connecting the free points of \textbf{Frame B} and \textbf{A} along which a $S_6 \ltimes (\mathbb{Z}_3^6)/\mathbb{Z}_3$ symmetry is preserved. This can be seen from the superconformal index by charging it with respect to the discrete symmetry.
At the free point of \textbf{Frame A} the symmetry enhanches to $SU(6)$ and $S_6$ becomes the Weyl group. 
It would be interesting to investigate this symmetry enhancement directly from \textbf{Frame B}. Indeed, even at the level of the index, the enhancement to $SU(6)$ is only visible thanks to the dual weakly coupled description provided by \textbf{Frame A}. Generally we do not expect that such a dual description is available in a point of the conformal manifold with enhanced symmetry, therefore an alternative approach to the study of this phenomenon is desirable.

 \section*{Acknowledgments}
 We thank Shlomo Razamat for useful comments during the preparation of this paper.
This work has been supported in part by the Italian Ministero dell’Istruzione, Università e Ricerca (MIUR), in part by Istituto Nazionale di Fisica Nucleare (INFN) through the “Gauge Theories, Strings, Supergravity” (GSS) research project and in
part by MIUR-PRIN contract 2017CC72MK-003.

\appendix

\section{Deconfining  $\mathcal{N}=4$ $SO(6)$ SYM}
In this appendix we reproduce some of the manipulations exploited throughout the main body of the paper, such as matching of discrete symmetries and symmetry enhancement on the conformal manifold, in the more familiar context of $\mathcal{N}=4$ SYM. Consider a $\mathcal{N}=1$ theory with one $SO(6)$ gauge group, three $USp(2)_i$ gauge groups and one bifundamental field $X_i$ between $SO(6)$ and each of the symplectic groups. The superpotential is:
\begin{equation}
W = X_1^2 \lbrace X_2^2, X_3^2 \rbrace
\end{equation}
The symplectic gauge groups confine and upon dualising them we obtain an $\mathcal{N}=1$ $SO(6)$ gauge theory with three adjoints $A_i$ and superpotential:
\begin{equation}
W= W^{\mathcal{N}=4} + \sum_{i=1}^3 Pf \left(A_i \right) = 
W^{\mathcal{N}=4} + \sum_{i=1}^3 Tr \left(A_i^3 \right)
\end{equation}
This is nothing but $\deform$-deformed $\mathcal{N}=4$ SYM \cite{Leigh:1995ep}. The $\deform$ deformation is exacly marginal and we can turn it off, moving on the conformal manifold, and recover $\mathcal{N}=4$ supersymmetry. In particular, from the $\mathcal{N}=1$ perspective, we recover the $SU(3)_R$ global symmetry. This $SU(3)_R$ is broken by the $\deform$ deformation to $S_3 \ltimes (\mathbb{Z}_3^3)/\mathbb{Z}_3$ where $S_3$ permutes the three adjoints and $\mathbb{Z}_3^{i}$ rotates $A_i$ by a phase $\frac{2}{3} \pi i$. 

The quiver theory has the same discrete symmetry, indeed there is a classical $S_3$ permuting the bifundamentals $X_i$ while $(\mathbb{Z}_3^3)/\mathbb{Z}_3$ comes from the anomalous “axial" symmetries as follows. Consider the $U(1)_i$ symmetries under which $X_i$ has charge $\frac{1}{2}$ and the other fields are not charged. These have an anomaly with one of the $USp(2)$ gauge groups and are broken to $\mathbb{Z}_6$, which are consistent with the superpotential. A $\mathbb{Z}_2$ subgroup of these symmetries is already contained in the $SO(6)$ gauge group, therefore they are broken to $\mathbb{Z}_3$. Anomaly cancellation with respect to $SO(6)$ further break them to $(\mathbb{Z}_3^3)/\mathbb{Z}_3$.

We can check the map between the global discrete symmetry by charging the superconformal index with $\mathbb{Z}_3$-valued fugacities. For $\mathcal{N}=4$ we have:
\begin{eqnarray}
\mathcal{I}^{\mathcal{N}=4} &=& 1 +  (pq)^{\frac{1}{3}} (p+q) \mathbf{3}_t + (pq)^{2\over3} \mathbf{6}_t + pq(1\!+\!\mathbf{10}_t \!-\! \mathbf{8}_t) 
\nonumber\\
+ \dots&=&
1 + (pq)^{\frac{1}{3}} (p+q) \mathbf{3}_t + (pq)^{2\over3} \mathbf{6}_t + (pq) 3 + \dots
\nonumber \\
\end{eqnarray}
where $t_i$, $i=1,2,3$ parametrize $SU(3)_R$ and the second equality is true after imposing $t_i^3=1$, consistently with the $\deform$-deformation. Similarly the superconformal index for the quiver theory is:
\begin{eqnarray}
\mathcal{I}^{\mathcal{N}=1}
&=&
1 + (pq)^{\frac{1}{3}} (p+q) \mathbf{3}_w+ (pq)^{2\over3} \mathbf{6}_w+ (pq) 3 + \dots
\nonumber\\
\end{eqnarray}
where $w_i$ are $\mathbb{Z}_3$-valued fugacities associated with $(\mathbb{Z}_3^3)/\mathbb{Z}_3$. The two indices match with $w_i = t_i$. We see that we are able to match the discrete $(\mathbb{Z}_3^3)/\mathbb{Z}_3$ symmetry that is preserved along the direction of the conformal manifold parametrized by the $\deform$-deformation. As a byproduct the $S_3$ symmetry is visible at the level of the index as well, and we can track it as we turn off the $\deform$-deformation until we reach the point in the conformal manifold where $\mathcal{N}=4$ is recovered. Here $S_3$ becomes the Weyl of the enhanced global symmetry $SU(3)_R$.

\bibliography{prd}

\end{document}